\documentclass[aps,prl,preprint,superscriptaddress]{revtex4-1}
\usepackage{graphicx}
\usepackage{dcolumn}
\usepackage{color}
\usepackage{textcomp}
\usepackage{multirow}
\usepackage{hyperref}
\usepackage[normalem]{ulem}
\usepackage{amsmath}
\setcitestyle{super}
\usepackage{soul}
\usepackage[dvipsnames]{xcolor}

\begin{document}

\title{Site-specific electronic and magnetic excitations of the skyrmion material Cu$_2$OSeO$_3$}

\author{Yanhong Gu}
\affiliation{National Synchrotron Light Source II, Brookhaven National Laboratory, Upton, NY 11973, USA.}
\author{Yilin Wang}     
\altaffiliation[Present address: ]{Hefei National Laboratory for Physical Sciences at Microscale, University of Science and Technology of China, Hefei, Anhui 230026, China.}
\affiliation{Department of Condensed Matter Physics and Materials Science, Brookhaven National Laboratory, Upton, New York 11973, USA.} 

\author{Jiaqi Lin}     
\altaffiliation[Present address: ]{School of Science, Westlake University, Hangzhou 310024, Zhejiang, China.}
\affiliation{Department of Condensed Matter Physics and Materials Science, Brookhaven National Laboratory, Upton, New York 11973, USA.}

\author{Jonathan Pelliciari}
\affiliation{National Synchrotron Light Source II, Brookhaven National Laboratory, Upton, NY 11973, USA.}

\author{Jiemin Li}
\affiliation{National Synchrotron Light Source II, Brookhaven National Laboratory, Upton, NY 11973, USA.}

\author{Myung-Geun Han}
\affiliation{Department of Condensed Matter Physics and Materials Science, Brookhaven National Laboratory, Upton, New York 11973, USA.}

\author{Marcus Peter Schmidt}
\affiliation{Max Planck Institute for Chemical Physics of Solids, N\"othnitzer Stra\ss e 40, D-01187 Dresden, Germany.}

\author{Gabriel Kotliar}     
\affiliation{Department of Condensed Matter Physics and Materials Science, Brookhaven National Laboratory, Upton, New York 11973, USA.} 
\affiliation{Department of Physics and Astronomy, Rutgers University, Piscataway, New Jersey 08856, USA.}

\author{Claudio Mazzoli}
\affiliation{National Synchrotron Light Source II, Brookhaven National Laboratory, Upton, NY 11973, USA.}
\author{Mark P. M. Dean}     
\affiliation{Department of Condensed Matter Physics and Materials Science, Brookhaven National Laboratory, Upton, New York 11973, USA.} 

\author{Valentina Bisogni}
\email[]{bisogni@bnl.gov}
\affiliation{National Synchrotron Light Source II, Brookhaven National Laboratory, Upton, NY 11973, USA.}

\begin{abstract}

\textbf{The manifestation of skyrmions in the Mott-insulator Cu$_2$OSeO$_3$ originates from a delicate balance between magnetic and electronic energy scales. As a result of these intertwined couplings, the two symmetry-inequivalent magnetic ions, Cu-I and Cu-II, bond into a spin S=1 entangled tetrahedron. 
However, conceptualizing the unconventional properties of this material and the energy of the competing interactions is a challenging task due the complexity of this system. Here we combine X-ray Absorption Spectroscopy and Resonant Inelastic X-ray Scattering to uncover the electronic and magnetic excitations of Cu$_2$OSeO$_3$ with site-specificity. We quantify the energies of the 3\textit{d} crystal-field splitting for both Cu-I and Cu-II, fundamental to optimize model Hamiltonians. Additionally, we unveil a site-specific magnetic mode, indicating that individual spin character is preserved within the entangled-tetrahedron picture. Our results thus provide experimental constraint for validating theories that describe the interactions of Cu$_2$OSeO$_3$, highlighting the site-selective capabilities of resonant spectroscopies.}
\end{abstract}
\maketitle
{\bf{INTRODUCTION}}\\
Since their first observation in magnetic solids \cite{muhlbauer_skyrmion_2009, yu_real-space_2010}, skyrmions -- nano-sized, topological spin objects -- immediately attracted enormous interest, 
thanks to their unique mobility properties in response to low current and electric fields \cite{nagaosa_topological_2013, finocchio_magnetic_2016, fert_magnetic_2017}. 
Skyrmions are consequently appealing for energy-efficient applications and their generation in insulators is furthermore attractive due  to reduced heat dissipation and fast switching response. Cu$_2$OSeO$_3$ is one of the few known Mott insulator hosting skyrmions \cite{seki_observation_2012,seki_formation_2012,ruff_magnetoelectric_2015,qian_new_2018}.
Normally, chiral, noncentrosymmetric, cubic magnetic materials are potential hosts for skyrmions \cite{lohani_quantum_2019}, but in the multiferroic Cu$_2$OSeO$_3$ the formation of such topological states furthermore arises from the 
delicate balance between the 
super-exchange couplings and the Dzyaloshinskii-Moriya (DM) interactions \cite{janson_quantum_2014, romhanyi_entangled_2014, ozerov_establishing_2014,zhang_magnonic_2020}. 

While Cu$_2$OSeO$_3$ shares the same $P2_13$ space group as the skyrmion-prototype MnSi \cite{muhlbauer_skyrmion_2009}, 
the fundamental magnetic unit behind the skyrmion nucleation within the ferrimagnetic phase is believed to be a composite Cu$_4$ tetrahedron with an effective spin S=1 \cite{janson_quantum_2014} 
and involving 
two differently coordinated Cu ions, Cu-I and Cu-II. 
Several studies \cite{janson_quantum_2014, romhanyi_entangled_2014, ozerov_establishing_2014, portnichenko_magnon_2016} identified the microscopic interactions between the individual ions within the magnetic building block as crucial to unravel the quantum nature of the skyrmions in Cu$_2$OSeO$_3$ and to explain the emergence of other unconventional phases \cite{white_electric-field-driven_2018, qian_new_2018,aqeel_microwave_2021}. 

As most experimental works focused so far on site-averaged magnetic properties of Cu$_2$OSeO$_3$ \cite{ozerov_establishing_2014, portnichenko_magnon_2016, tucker_spin_2016, versteeg_inelastic_2019}, the site-specific magnetic response and the local electronic structure of the Cu 3\textit{d} valence states remain marginally understood \cite{langner_coupled_2014, versteeg_optically_2016}. 
Nonetheless, the latter is fundamentally related to the DM interactions -- responsible for the helical order and the skyrmion formation \cite{yang_strong_2012} -- as they are explained through multi-orbital and multi-site hopping paths, involving both Cu-I and Cu-II. 
\textcolor{black}{On the other hand, the site-specific magnetic response can elucidate the validity of the Cu$_4$ tetrahedron picture at the microscopic level, exploiting the single spin point of view.}

Hence, the complex nature of Cu$_2$OSeO$_3$ calls for site-specific magnetic and electronic investigation\textcolor{black}{s. These results will build prerequisite information for extracting} all the interactions underlying the skyrmion generation in this system \textcolor{black}{-- by experimentally validating microscopic theoretical models --} ultimately unveiling the \textcolor{black}{real} energy balance that stabilizes the skyrmion phase. Such information is crucial for future skyrmion applications, as it promises the possibility for designing optimized materials -- i.e. with an extended skyrmion pocket in the magnetic phase diagram -- where the key-interactions can be tuned either by film thickness, electric field, pressure, and strain \cite{seki_observation_2012, ruff_magnetoelectric_2015, deng_room-temperature_2020, burn_field_2020}.

Here, we combine X-ray Absorption Spectroscopy (XAS) and Resonant Inelastic X-ray Scattering (RIXS) to unveil the 
electronic and magnetic excitations 
associated with the two inequivalent Cu sites in the ferrimagnetic phase of Cu$_2$OSeO$_3$ ($T_\text{C} \simeq$ 57 K).  Using density functional theory (DFT) and RIXS cross section calculations, we disentangled the resonant energies associated with Cu-I and Cu-II ions. Capitalizing on this finding, we determine the site-specific spectral fingerprints and extract the orbital symmetries and energies of the crystal-field split 3$\textit{d}$ levels for both Cu-I and Cu-II. 
Furthermore, we reveal a site-specificity of the medium-energy magnon mode around 35 meV. 
Our results thus provide an experimental constraint for theories aimed at quantifying the competing energy terms underlying the skyrmion formation. More broadly, this approach can be extended to the design of devices and heterostructures with improved skyrmion properties, highlighting the importance of resonant techniques when dealing with multi-site complex systems.

\begin{figure}
\includegraphics[width=0.5\textwidth]{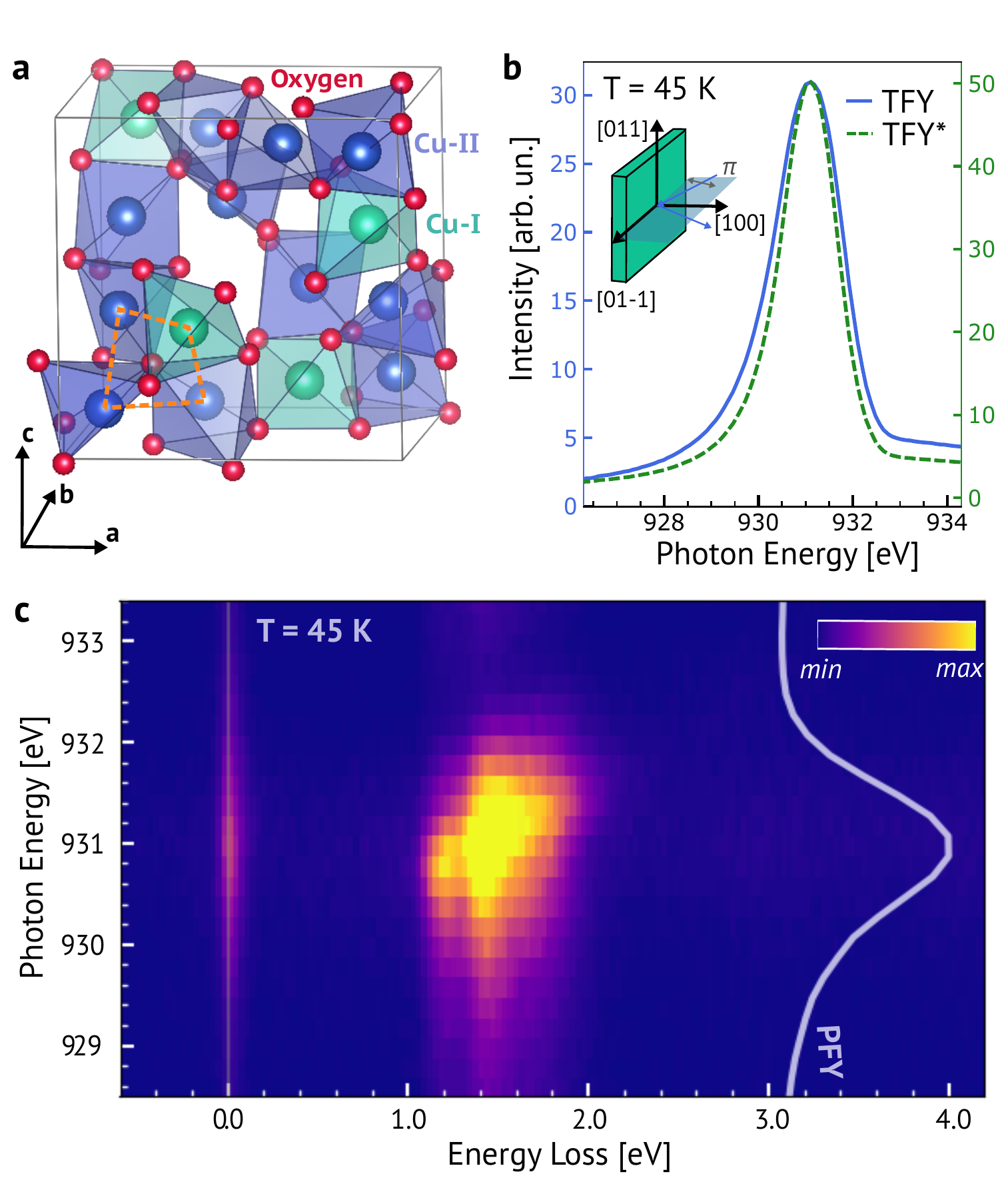}
\caption{{\bf Overview of the crystal structure, experimental configuration, and Cu L$_3$ XAS and RIXS data.} {\bf a} Cu$_2$OSeO$_3$ unit cell ($a = 8.925$ \AA). The orange dashed line highlights the magnetic tetrahedron unit. {\bf b} Cu L$_3$ XAS spectrum \textcolor{black}{in terms of TFY}, at $T$ = 45 K. \textcolor{black}{ TFY$^*$ is the spectrum after correction for self-absorption and saturation effects.} {\bf c} RIXS energy map with linear color scale for the intensity, also at $T$ = 45 K. The thin solid line displays the partial fluorescence yield (PFY) signal, obtained integrating the RIXS spectra up to 10 eV. The inset in panel {\bf b} displays the sample orientation used for these measurements, with [100] and [01-1] axes lying within the scattering plane. Incoming $\pi$ polarized x-rays were used for all measurements.
}
\label{fig:fig1}
\end{figure}

{\bf{RESULTS}}\\
{\bf{Experimental details --}} Single crystalline Cu$_2$OSeO$_3$ was prepared by chemical vapor transport method \cite{portnichenko_magnon_2016, qian_new_2018}. Figure \ref{fig:fig1}{\bf a} shows the crystallographic unit cell of Cu$_2$OSeO$_3$, containing 16 Cu atoms arranged in four tetrahedrons. Each tetrahedron (dashed orange line in Fig. \ref{fig:fig1}{\bf a}) consists of one Cu-I and three Cu-II ions. Below the ferrimagnetic ordering temperature $T_\text{C}$, the Cu-I spin aligns anti-parallel to the Cu-II spins. As each Cu brings a spin momentum of 1/2, this configuration yields a total spin momentum of S=1 for the tetrahedron unit. A Curie temperature of $T_\text{C} \simeq$ 57 K was extracted from the sample used in this study, in line with Ref. [\citenum{wu_physical_2015}]. For the XAS and RIXS measurements, we prepared a single crystal with [100] surface normal. The sample orientation used throughout the experiment is displayed in the inset of Fig. \ref{fig:fig1}{\bf b}. 

{\bf Soft X-ray Resonant Spectroscopies of Cu$_2$OSeO$_3$ --} Resonant spectroscopies with their chemical sensitivity provide unique advantages in the study of multi-site compounds. 
RIXS is furthermore helpful when there is need to study the valence electronic structure as it probes the charge-neutral, dipole-forbidden, \textit{dd}-excitations of a system, enabling to reconstruct its ground-state energy levels \cite{ament_resonant_2011, sala_energy_2011, bisogni_orbital_2015, elnaggar_magnetic_2019, lebert_resonant_2020, occhialini_local_2021}. 
In particular, for 3\textit{d} elements, such excitations can be accessed using the L$_3$ edge resonance in the soft x-ray range, promoting electrons from the $2\textit{p}_{3/2}$ core states to the 3\textit{d} valence states. 

Figure~\ref{fig:fig1}{\bf b} presents the Cu L$_3$ XAS spectrum of Cu$_2$OSeO$_3$ \textcolor{black}{(blue line) acquired in terms of Total Fluorescence Yield (TFY),} at $T = 45$~K. The line-shape and the peak at 930.9 eV are consistent with previous measurements \cite{zhang_resonant_2016}, although our interpretation differs from Ref. [\citenum{langner_coupled_2014}] as explained later on in the text. 
Since a single Cu 3$d^9$ site hosts one hole, its XAS spectrum is expected to be a Lorentzian curve with a $2p$ core-hole lifetime-dominated width of 0.3-0.5 eV \cite{krause_natural_1979, bisogni_femtosecond_2014}. \textcolor{black}{While the enhanced width of $\sim 1.2$~eV and the asymmetric line-shape suggest distinct contributions from the Cu-I and Cu-II sites, we underline that their energy splitting $\Delta \text E_{\text{ CuII-CuI}}$ should be intrinsically small. This is supported by TFY$^*$ (green dashed line), corrected TFY for self-absorption and saturation effects \cite{haskel1999fluo,XAS_corr}, that still displays an asymmetric single-peaked line-shape with $\sim 1$~eV large width.} 


No clear consensus has been reached so far on the value of $\Delta \text E_{\text{ CuII-CuI}}$. Previous DFT+U calculations estimated $\Delta \text E_{\text{ CuII-CuI}}$ to be $\sim$0.2 eV \cite{yang_strong_2012, versteeg_optically_2016}, while resonant x-ray scattering proposed $\Delta \text E_{\text{ CuII-CuI}}$ to be \textcolor{black}{$\sim$2 eV} \cite{langner_coupled_2014}. Here, we quantify $\Delta \text E_{\text{ CuII-CuI}}$ 
 and resolve the 3\textit{d} electronic structure for each Cu site using RIXS at the Cu L$_3$ edge. The spectra acquired by varying the incident photon energy within the range $\sim$ 928 eV - 933 eV are plotted versus energy loss  
and gathered in a color map (see Fig.~\ref{fig:fig1}{\bf c}). 
This displays two main features: the quasi-elastic line around zero energy loss and a broad multi-peaked structure between 1 eV and 2 eV. The latter excitations are interpreted as intra-site \textit{dd}-excitations stemming from the local crystal field in agreement with ellipsometry data \cite{versteeg_optically_2016} and other RIXS measurements on Cu$^{2+}$ systems \cite{sala_energy_2011,bisogni_orbital_2015}.

\begin{figure}
\includegraphics[width=0.5\textwidth]{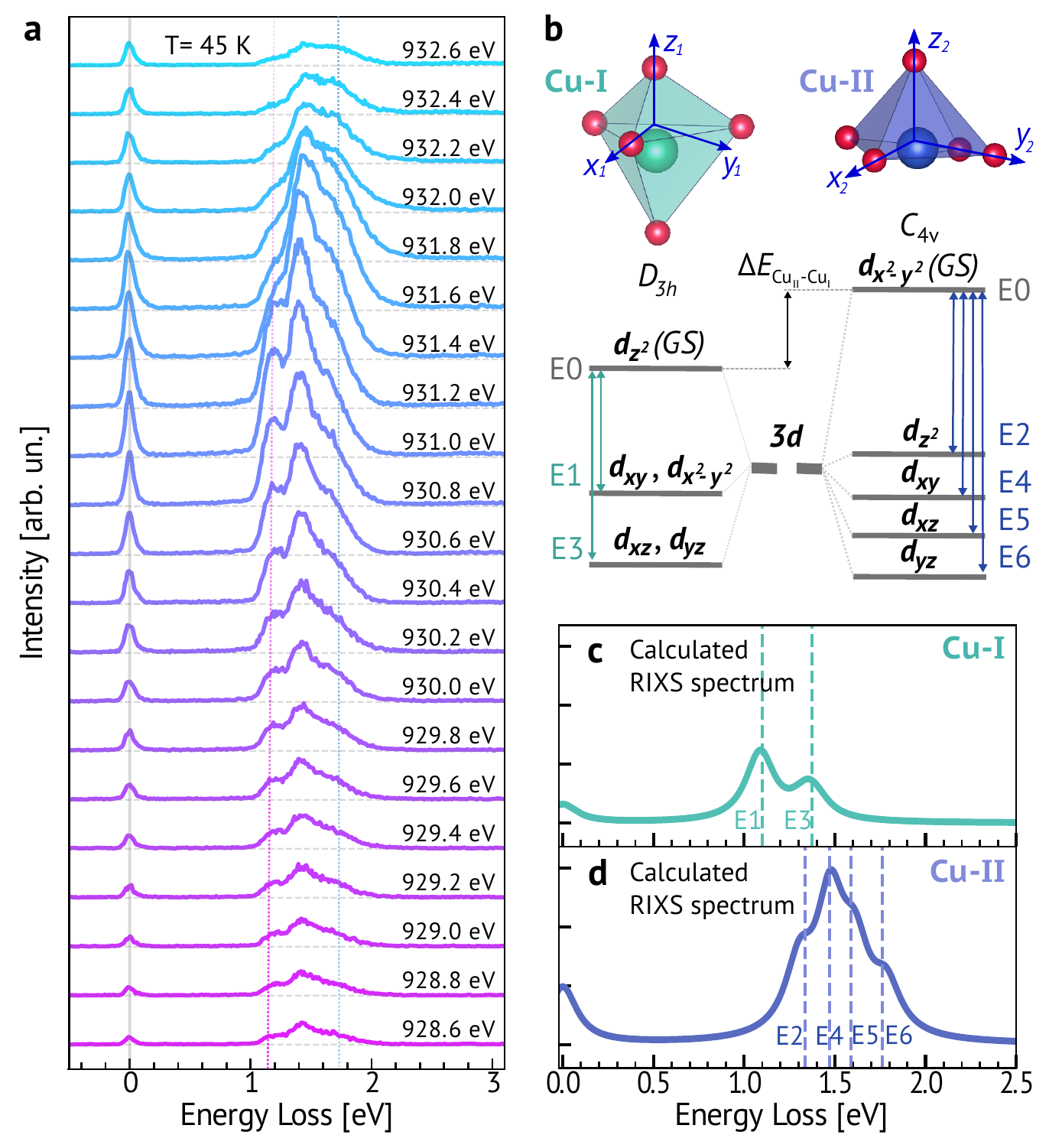}
\caption{{\bf \textit{dd}-excitations of Cu-I and Cu-II.} 
{\bf a} RIXS spectra as a function of incident photon energy, for a selected incident energy range: 928.6 eV to 932.6 eV. {\bf b} Crystal-field splittings of the Cu-I (Cu-II) 3\textit{d} orbitals in $D_{3h}$ ($C_{4v}$) point group symmetry, respectively. The 3\textit{d} orbitals are defined with respect to the local Cartesian coordinates indicated in blue. 
The vertical arrows labeled E1 to E5 associate the \textit{dd}-excitations with the respective initial and final 3\textit{d} levels. {\bf c, d} Calculated RIXS responses for the individual Cu-I and Cu-II ions.
}
\label{fig:fig2}
\end{figure}

The fine details of the \textit{dd}-excitations are shown in 
Fig. \ref{fig:fig2}{\bf a}, where the RIXS spectra are presented as a vertical stack for increasing incident photon energy. 
While the overall intensity evolution of the spectral weight is due to changing the photon energy across the absorption resonance, the shape evolution between 930.6 eV and 932 eV highlights the existence of different sets of crystal-field excitations. 
The excitations pattern originates from the presence of the two inequivalent Cu species, Cu-I and Cu-II, having different oxygen coordination.  
However, because of the small splitting 
between Cu-I and Cu-II energy levels as suggested by the XAS, it is necessary to resort to a more involved data analysis for disentangling the site-specific crystal field excitations and their resonant energies. 

\begin{figure*}
\includegraphics[width=0.95\textwidth]{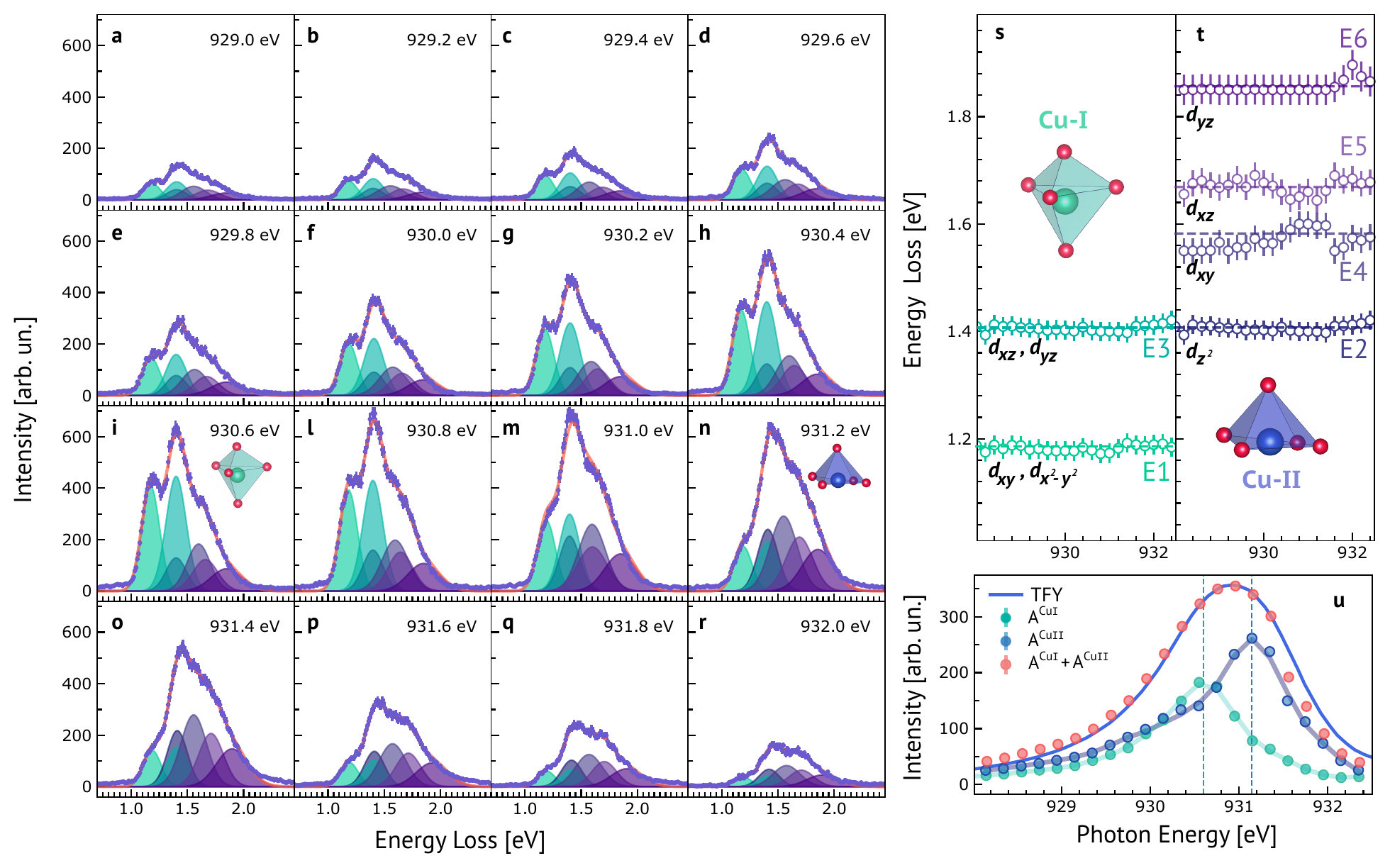}
\caption{{\bf Analysis of \textit{dd}-excitations in Cu$_2$OSeO$_3$.} {\bf a-r} RIXS spectra (dots), in the 0.8-2.5 eV region, as a function of incident energy across the Cu $L_3$ edge. \textcolor{black}{The error bars are defined assuming a Poisson distribution of the single-photon counted events.} The fitted components are colored in aquamarine for the Cu-I \textit{dd}-excitations (E1 and E3), and in \textcolor{black}{purple} for the Cu-II \textit{dd}-excitations (E2, E4, E5 and E6). The sum of the fitted components is represented as a pink, thick solid line. {\bf s-t} Summary of the fitted \textit{dd}-excitation energies across the Cu $L_3$ edge, with assigned orbital character following DFT calculations. {\bf u} Site-resolved \textit{dd}-excitation amplitudes A$^{\text{CuI}}$ (aquamarine dots) and A$^{\text{CuII}}$ (blue dots). \textcolor{black}{The error bars are contained within the marker size}. The smoothed lines underneath are a guide for the eyes. Their sum is displayed by the pink dots. A rescaled \textcolor{black}{TFY} profile is plotted in the same figure using a blue solid line. Vertical dotted lines mark the resonant energy extracted for Cu-I ion at 930.6 eV and Cu-II ion at 931.15 eV.
}
\label{fig:fig3}
\end{figure*}

{\bf{DISCUSSION}}\\
{\bf Resolving Cu-I and Cu-II electronic structure using DFT and single-ion calculations --}
Preliminary considerations on the orbital  excitations of the two sites can be made by examining the distinct point group symmetries of the crystal field associated with Cu-I, approximately a trigonal bypiramid $D_{3h}$, and Cu-II, approximately a square pyramid $C_{4v}$ \cite{bos_magnetoelectric_2008, versteeg_optically_2016}. 
These two symmetries naturally lead to different orbital arrangements \cite{versteeg_optically_2016}. The ground state (GS) orbitals can be identified by referring to the the local Cartesian axis (see Fig.~\ref{fig:fig2}{\bf b}) that minimizes the Cu-O distance ($z_1$ for Cu-I yielding a $d_{z^2}$ GS, and $x_2/y_2$ for Cu-II yielding a $d_{x^2-y^2}$ GS), in agreement with Refs. [\citenum{yang_strong_2012,versteeg_optically_2016}]. 
To obtain an estimate of the site-specific orbital character and energy of the crystal field split 3\textit{d} levels in Cu$_2$OSeO$_3$, we perform DFT combined with Wannier90 calculations. Details on the DFT part can be found in the Supplementary Information, Sec. I. Tight binding (TB) Hamiltonians $\hat{H}^{\text{CF}}_{\text{Cu-I}}$ and $\hat{H}^{\text{CF}}_{\text{Cu-II}}$ consisting of 3\textit{d} orbitals from four Cu-I sites and from twelve Cu-II sites are then formulated. 
From the calculated 3\textit{d} energies, we can directly extract the crystal field excitation energies for each Cu site and use them as a guide to fit the experimental data. Together with the schematics of Fig.~\ref{fig:fig2}{\bf b}, we thus expect two \textit{dd}-excitations for Cu-I, labeled as E1 and E3, and four \textit{dd}-excitations for Cu-II, labeled as E2, E4, E5 and E6. Note that E0 corresponds to the zero energy transition, obtained when the initial and final 3\textit{d} levels coincide. 

To address the Cu-I and Cu-II resonant energies as well as their respective orbital energies, we simulate XAS and RIXS spectra based on single-atom model using the EDRIXS code~\cite{wang_edrixs_2019}. Details are collected in the Supplementary Information, Sec. II.  
The intensity of the calculated RIXS spectra accounts for all atoms within the unit cell, neglecting interference effects between them. Furthermore, the experimental geometry and the incoming polarization projections are included as well. By using the theoretical energies for the \textit{dd}-excitations (see Tab.~S3 in the Supplementary Information) and the RIXS intensity obtained from the atomic model calculations, we can simulate the RIXS spectra associated with each Cu species, see Fig.~\ref{fig:fig2}{\bf c} for Cu-I and Fig.~\ref{fig:fig2}{\bf d} for Cu-II. 
We performed then a constrained fit where the integrated intensity of each $dd$ transition is fixed to the theoretical model while their energy and width are allowed to vary. As detailed in the Supplementary Information, Sec. III, we use the following model to fit the data:
\begin{equation}
\label{eqn:fitting_model}
I_\text{RIXS}^\text{total}= A^{\text{CuI}} \cdot \tilde{I}_\text{RIXS}^{\text{CuI}} + A^{\text{CuII}} \cdot \tilde{I}_\text{RIXS}^{\text{CuII}}
\end{equation}
where A$^{\text{CuI/II}}$ are the site-specific amplitudes at each incident energy and $\tilde{I}_\text{RIXS}^{\text {CuI/II}}$ are the  site-specific area-normalized theoretical RIXS spectra.

\begin{table}
    \centering
    \begin{tabular}{c c c }
        \hline
        {\bf{    Site    }} & {\bf{  Excitation  }} & {\bf{  Energy [eV]  }} \\[.5ex]
        \hline
        \multirow{2}{*}{{\bf Cu-I}}  & \multicolumn{1}{c}{E1} &  \multicolumn{1}{c}{1.18 $\pm$ 0.01}\\
        & \multicolumn{1}{c}{E3} & \multicolumn{1}{c}{1.40 $\pm$ 0.01}  \\
        \hline
        \multirow{4}{*}{{\bf Cu-II}}  & \multicolumn{1}{c}{E2} &  \multicolumn{1}{c}{1.4 $\pm$ 0.01} \\
        & \multicolumn{1}{c}{E4} & \multicolumn{1}{c}{1.57 $\pm$ 0.01} \\
        & \multicolumn{1}{c}{E5} & \multicolumn{1}{c}{1.67 $\pm$ 0.01} \\
        & \multicolumn{1}{c}{E6} & \multicolumn{1}{c}{1.86 $\pm$ 0.01} \\
        \hline
    \end{tabular}
    \caption{Energies of the Cu-I and Cu-II $dd$-excitations, extracted from the fit of the RIXS measurements. The errors are defined as the standard deviation associated with the least square fit results.}
    \label{tab:tab1}
\end{table}

\begin{figure}
\includegraphics[width=0.5\textwidth]{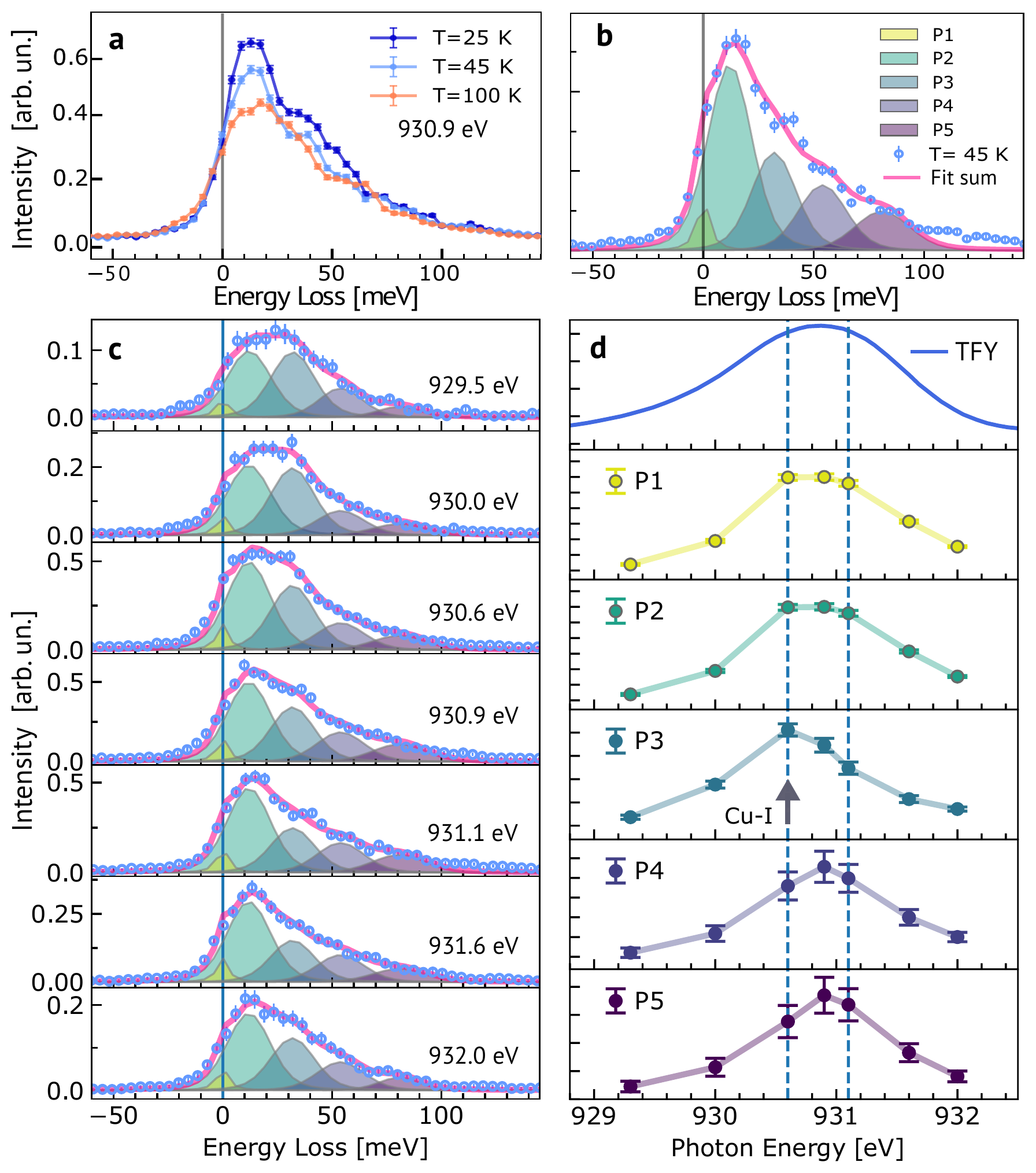}
\caption{{\bf Magnetic excitations in Cu$_2$OSeO$_3$.} {\bf a} High-resolution RIXS spectra measured at an incident energy of 930.9 eV, $q$ = [1.3,0,0] r.l.u. for $T$ = 27 K, 45 K, and 100 K, with $\pi$ polarized light. The zero energy was determined with reference to a carbon tape placed on the sample. {\bf b} Fitting analysis of the 45 K RIXS spectrum, using five Gaussian peaks with fixed width, fixed center position (allowing $\pm 5\%$ variation w.r.t. the values presented in the main text) and free amplitude. Raw data are displayed as open dots. Fitting sum is the solid line. \textcolor{black}{The error bars are defined as in Fig.~\ref{fig:fig3}}. {\bf c} Repeating the fitting analysis for all RIXS spectra measured across the Cu L$_3$ resonance. {\bf d} Summary of the integrated intensity for each fitting component, as a function of the incident photon energy (dot symbols). \textcolor{black}{The error bars are extracted from the fitting, through the error propagation method}. The smoothed lines underneath are a guide for the eyes. The \textcolor{black}{Cu L$_3$ absorption} spectrum is reproduced on top to ease the comparison.
}
\label{fig:fig4}
\end{figure}

The fitted spectra are presented in Fig.~\ref{fig:fig3}{\bf a-r}, where each panel refers to a specific incident energy and displays the raw data (open dots), the fitted Gaussian components for each \textit{dd}-excitations (solid filling in aquamarine color for Cu-I and \textcolor{black}{purple} color for Cu-II) and their sum (pink solid line). Figure~\ref{fig:fig3}{\bf s-t} summarizes the fitted \textit{dd}-excitation center positions, corresponding to the E1-E6 values. Notably, the extracted peak positions are reasonably constant across the scanned incident energy range, validating the reliability of the fitting model, while their averages and errors are summarized in the fourth column of Tab.~\ref{tab:tab1}. 

From these results, we conclude that the experimental \textit{dd}-excitation energies are well reproduced by the eigenvalues of the $\hat{H}^{\text{CF}}_{\text{Cu-I}}$ and $\hat{H}^{\text{CF}}_{\text{Cu-II}}$ Hamiltonians (within 10-20\%), and confirm that the effective $D_{3h}$ symmetry for Cu-I ion and the distorted $C_{4v}$ symmetry for Cu-II ion are good approximations for the real material despite small distortions from these idealized symmetries. 

Furthermore, by plotting the A$^{\text{CuI}}$ and A$^{\text{CuII}}$ amplitudes as a function of the incident photon energy, we obtain the experimental resonant profiles for Cu-I and Cu-II sites (see Fig.~\ref{fig:fig3}{\bf u}). By summing the A$^{\text{CuI}}$ and A$^{\text{CuII}}$ amplitudes across the Cu L$_3$ edge (pink dots in Fig.~\ref{fig:fig3}{\bf u}) we can well reproduce the XAS of Cu$_2$OSeO$_3$ (blue solid line): this good agreement corroborates the consistency of the analysis. After extracting the respective maxima position from the A$^{\text{CuI}}$ ($\sim$ 930.6 eV) and A$^{\text{CuII}}$ ($\sim$ 931.15 eV) profiles, we can additionally quantify $\Delta E_{\text{CuII-CuI}}\sim$0.55 $\pm$ 0.05 eV, as the energy difference between the Cu-I and Cu-II resonances. This value is a bit larger than our theoretical estimate ($\sim$ 0.33 eV, refer to Sec. I of Supplementary Information) and previous DFT+U work \cite{versteeg_optically_2016, yang_strong_2012}, while it strongly differs from the value reported in Ref. [\citenum{langner_coupled_2014}], \textcolor{black}{$\sim$2 eV}. Our result finally legitimates the approximation of neglecting interference-effects between the two inequivalent Cu species, since $\Delta \text E_{\text{ CuII-CuI}}\gtrsim \Gamma_{CuL}$.

{\bf Site-dependent magnetic excitations --} \textcolor{black}{The magnetic properties of Cu$_2$OSeO$_3$ as well as the magnon modes have been explained so far as emanating from the effective S=1 Cu$_4$ tetrahedra \cite{janson_quantum_2014,romhanyi_entangled_2014,portnichenko_magnon_2016,tucker_spin_2016,ozerov_establishing_2014}, rather than from individual Cu spins, yielding the definition of ``entangled tetrahedron ground state''. However, some discrepancies between the Cu$_4$ tetrahedra model and the spin excitations have been reported in Ref. [\citenum{neutron_Tucker2016}]. Benefiting from the unique site-sensitivity offered by RIXS as demonstrated above, we investigate the site-dependence of the magnon modes in Cu$_2$OSeO$_3$, to assess possible contributions of the two inequivalent Cu sites into the spin excitations improving our current understanding of this complex system.}


The magnon spectrum of Cu$_2$OSeO$_3$ has been studied so far by several techniques, e.g. inelastic neutron scattering, Raman, Infrared, ESR and THz spectroscopy  \cite{portnichenko_magnon_2016,tucker_spin_2016,gnezdilov_magnetoelectricity_2010, miller_magnetodielectric_2010,ozerov_establishing_2014}. These results consistently reported $i)$ inter-tetrahedron ferromagnons below 15 meV; $ii)$ intra-tetrahedron, medium energy magnon branches between 30-40 meV; $iii)$ high-energy ($>$ 50 meV) phonon modes and multi-magnons. In Fig.~\ref{fig:fig4} we display high-resolution RIXS spectra of Cu$_2$OSeO$_3$ at $q$=[1.3,0,0] r.l.u., measured at the Cu L$_3$ edge. From Fig.~\ref{fig:fig4}{\bf a}, a long tail up to 100 meV can be observed. 
With reference to the magnon dispersion measured by inelastic neutron scattering \cite{portnichenko_magnon_2016,tucker_spin_2016}, at $q$=[1.3,0,0] r.l.u., we expect a first component around 12 meV from the ferromagnon mode, a second one around 35 meV from the medium-energy magnon branches. A phonon mode around  54 meV is furthermore expected, consistently with Raman and Infrared data \cite{gnezdilov_magnetoelectricity_2010, miller_magnetodielectric_2010}. Moreover, we attribute the high energy spectral-weight around 70-80 meV to multi-magnons. This assignment is further supported by the temperature dependence of Fig. \ref{fig:fig4}{\bf a} \cite{ellis_magnetic_2010}, where the spectral-weight of the magnetic components is enhanced below $T_\text{C}$\textcolor{black}{, up to 100~meV}. Using resolution limited Gaussian peaks (FWHM = 30 meV) for fitting the elastic peak (P1, 0 meV), the ferromagnon (P2, 12 meV), the medium-energy magnon (P3, 35 meV), the phonon (P4, 54 meV) and a wider Gaussian for  high-energy multi-magnon component (P5, 80 meV, FWHM = 40 meV), we can accurately reproduce the RIXS spectrum at $T$ = 45 K, see Fig.~\ref{fig:fig4}{\bf b}.

To investigate the site-dependent character of the magnetic excitation, we collected RIXS spectra as a function of incident photon energy across the Cu L$_3$ edge, while leaving $q$ unchanged. The resulting data are presented in Fig.~\ref{fig:fig4}{\bf c}. As expected, an overall amplitude renormalization of the whole spectrum takes place due to the absorption effect. However, by tracking the individual P1-P5 components through the fitting analysis, we can extract the intensity behaviour for each individual excitation as a function of the photon energy, see Fig.~\ref{fig:fig4}{\bf d}. 
\textcolor{black}{The intensity profile of the elastic peak P1, the ferromagnon mode P2, the phonon mode P4 and the multi-magnon mode P5 track well with the TFY absorption profile (peaked at $\sim$ 903.9 eV), within the error-bars. This result suggests these modes do not have any specific or noticeable Cu-site dependence. Interestingly, instead, the intensity profile of the medium-energy magnon mode P3 displays a pronounced resonance at the Cu-I sites, $\sim$ 930.6 eV, clearly standing out beyond the error-bar scale.}
We interpret this peculiar behaviour considering the 35 meV component of the RIXS spectra dominated by magnon modes with B and C character \cite{ozerov_establishing_2014, romhanyi_entangled_2014}: these intra-tetrahedron modes correspond to rotating the Cu-I minority spins through the $J^{AF}_S$ interaction, while leaving the Cu-II spins unaltered. Hence, our result highlights that individual Cu spin character persists in the medium-energy magnon modes, simultaneously with the entangled-tetrahedron nature reported so far for the magnetic excitations of Cu$_2$OSeO$_3$ \cite{romhanyi_entangled_2014,janson_quantum_2014}. This finding recalls the magnetic dual nature proposed for MnSi \cite{yaouanc_dual_2020}, thus highlighting the complexity of these systems.

{\bf Conclusion --} 
By combining resonant spectroscopies, DFT and single-ion calculations, we elucidated the electronic and magnetic excitations of the multi-site skyrmion material Cu$_2$OSeO$_3$. We identified the $L_3$ resonant energies for Cu-I and Cu-II ionic species present in this system. With this unique information at hand, we revealed the site-resolved 3\textit{d} electronic structure in terms of crystal field splittings, and moreover, the difference between the Cu-I and Cu-II ground state energies of about 0.55 eV. Due to the difficulty in accurately calculating these quantities for strongly correlated electron systems, our work provides \textcolor{black}{theorists with} an experimental benchmark for fine-tuning microscopic models of Cu$_2$OSeO$_3$, hence for extrapolating the competing energy terms, i.e. hopping, exchange integrals and DM interactions. 
Furthermore, we revealed an unexpected site-dependent character (Cu-I) for the medium-energy magnon branch: this result  demonstrates that individual spin character is preserved in specific magnon modes, suggesting that local spin behaviour coexists with the entangled nature of the magnetic ground state\textcolor{black}{, explained by means of S=1 tetrahedra rather than single spins}. 

\textcolor{black}{More broadly, as the magnetism in Cu$_2$OSeO$_3$ is determined by the competition between the super-exchange interactions, DM interactions and crystal-anisotropy stemming from the spin-orbital interactions, our results on site-specific magnetic and electronic ground state excitations should be regarded as a prerequisite for validating future and past theoretical models of Cu$_2$OSeO$_3$ dedicated to the microscopic understanding of i.e. the skyrmion phase, the magnetic chirality, and the multiferroicity.}   

\textcolor{black}{Finally, our} finding overall highlights the complexity of this skyrmion material and, at the same time, the relevance of using advanced spectroscopies to reveal site-specific information. 
\textcolor{black}{T}he method presented here can be extended to thin films and heterostructures \textcolor{black}{of Cu$_2$OSeO$_3$} as well as devices (e.g. under the application of electric field) to elucidate the evolution of the site-specific excitations, and thus of the energy balance between the interactions contributing to the skyrmion formation.

{\bf{METHODS}}\\
{\bf XAS and RIXS Measurements}. The XAS and RIXS experiments were performed at the SIX 2-ID beamline of NSLS-II \cite{dvorak_towards_2016}. The XAS data of Fig. \ref{fig:fig1}{\bf b} was measured in Total Fluorescence Yield (TFY), at an incident angle of $\theta_\text{in}=$ 20$^{\circ}$. The energy resolution and experimental geometry used for the RIXS measurements were: $\Delta$E=50 meV (FWHM) and $\theta_\text{in}=$ 20$^{\circ}$ / $2\Theta$=90$^{\circ}$ for the crystal field study; $\Delta$E=30 meV (FWHM) and $\theta_\text{in}=$ 75$^{\circ}$ / $2\Theta$=150$^{\circ}$ for the spin excitation study. All the measurements used $\pi$-polarized x-ray photons.

{\bf Calculations}\\
Details about the DFT and single-ion calculations are available respectively at Sec. I and Sec. II of the Supplementary Information. 

{\bf DATA AVAILABILITY}\\
Data that support the findings of this study are available upon reasonable request from the corresponding authors.

{\bf ACKNOWLEDGEMENTS}\\
This work was supported by the U.S. Department of Energy (DOE), Office of Science, Basic Energy Sciences, Early Career Award Program. Y.W. and G.K. were supported by the US Department of energy, Office of Science, Basic Energy Sciences as a part of the Computational Materials Science Program through the Center for Computational Design of Functional Strongly Correlated Materials and Theoretical Spectroscopy. This research used beamline 2-ID of the National Synchrotron Light Source II, a U.S. Department of Energy (DOE) Office of Science User Facility operated for the DOE Office of Science by Brookhaven National Laboratory under Contract No. DE-SC0012704.

{\bf{AUTHOR CONTRIBUTIONS}}\\
V.B. conceived the project, with input from M.P.M.D, M.-G.H., and C.M.; M.P.S. grew and characterized the sample. V.B., M.P.M.D., Y.G., J.Lin, J. Li, and J.P. performed the XAS and RIXS experiments. V.B. analysed and interpreted the data with the help of M.P.M.D., Y.G., J.Lin, C.M., and J.P;  Y.W. and M.P.M.D. performed the theory calculations, with help from G.K.; V.B. and Y.G. wrote the manuscript with input from all the authors.
\\{\bf{CORRESPONDENCE}}\\
Correspondence and requests for materials should be addressed to V. Bisogni.
\\{\bf{ADDITIONAL INFORMATION}}\\
{\bf Competing Interests:} The authors declare no competing interests.
\\{\bf Supplementary Information} is available for this paper at ...\\

\makeatletter
\renewcommand\@biblabel[1]{#1.}
\makeatother

\bibliographystyle{naturemag}
\bibliography{apssamp}

\end{document}


\title{Supplementary Information for: \\ Site-specific electronic and magnetic excitations of the skyrmion material Cu$_2$OSeO$_3$}

\author{Yanhong Gu}
\affiliation{National Synchrotron Light Source II, Brookhaven National Laboratory, Upton, NY 11973, USA.}
\author{Yilin Wang}     
\altaffiliation[Present address: ]{Hefei National Laboratory for Physical Sciences at Microscale, University of Science and Technology of China, Hefei, Anhui 230026, China.}
\affiliation{Department of Condensed Matter Physics and Materials Science, Brookhaven National Laboratory, Upton, New York 11973, USA.} 

\author{Jiaqi Lin}     
\altaffiliation[Present address: ]{Department of Physics, Westlake University, Hangzhou, China.}
\affiliation{Department of Condensed Matter Physics and Materials Science, Brookhaven National Laboratory, Upton, New York 11973, USA.}

\author{Jonathan Pelliciari}
\affiliation{National Synchrotron Light Source II, Brookhaven National Laboratory, Upton, NY 11973, USA.}

\author{Jiemin Li}
\affiliation{National Synchrotron Light Source II, Brookhaven National Laboratory, Upton, NY 11973, USA.}

\author{Myung-Geun Han}
\affiliation{Department of Condensed Matter Physics and Materials Science, Brookhaven National Laboratory, Upton, New York 11973, USA.}

\author{Marcus Peter Schmidt}
\affiliation{Max Planck Institute for Chemical Physics of Solids, N\"othnitzer Stra\ss e 40, D-01187 Dresden, Germany.}

\author{Gabriel Kotliar}     
\affiliation{Department of Condensed Matter Physics and Materials Science, Brookhaven National Laboratory, Upton, New York 11973, USA.} 
\affiliation{Department of Physics and Astronomy, Rutgers University, Piscataway, New Jersey 08856, USA.}

\author{Claudio Mazzoli}
\affiliation{National Synchrotron Light Source II, Brookhaven National Laboratory, Upton, NY 11973, USA.}
\author{Mark P. M. Dean}     
\affiliation{Department of Condensed Matter Physics and Materials Science, Brookhaven National Laboratory, Upton, New York 11973, USA.} 

\author{Valentina Bisogni}
\email[]{bisogni@bnl.gov}
\affiliation{National Synchrotron Light Source II, Brookhaven National Laboratory, Upton, NY 11973, USA.}

\maketitle

\section{Density Functional Theory Calculations}
We performed density functional theory (DFT) combined with Wannier90 calculations to estimate the crystal field splitting. The DFT part of the calculation has been done by the Vienna Ab-initio Simulation Package (VASP)~\cite{kresse_efficient_1996} with projector augmented-wave (PAW)
pseudopotential~\cite{blochl_projector_1994, kresse_ultrasoft_1999} and Perdew-Burke-Ernzerhof parametrization of the generalized gradient approximation (GGA-PBE)
exchange-correlation functionals~\cite{perdew_generalized_1996}. 
The energy cutoff of the plane-wave basis is set to be 500 eV, and a $\Gamma$-centered
$11 \times 11 \times 11$ $K$-point grid was used.

\begin{figure}[b]
\includegraphics[width=0.45\textwidth]{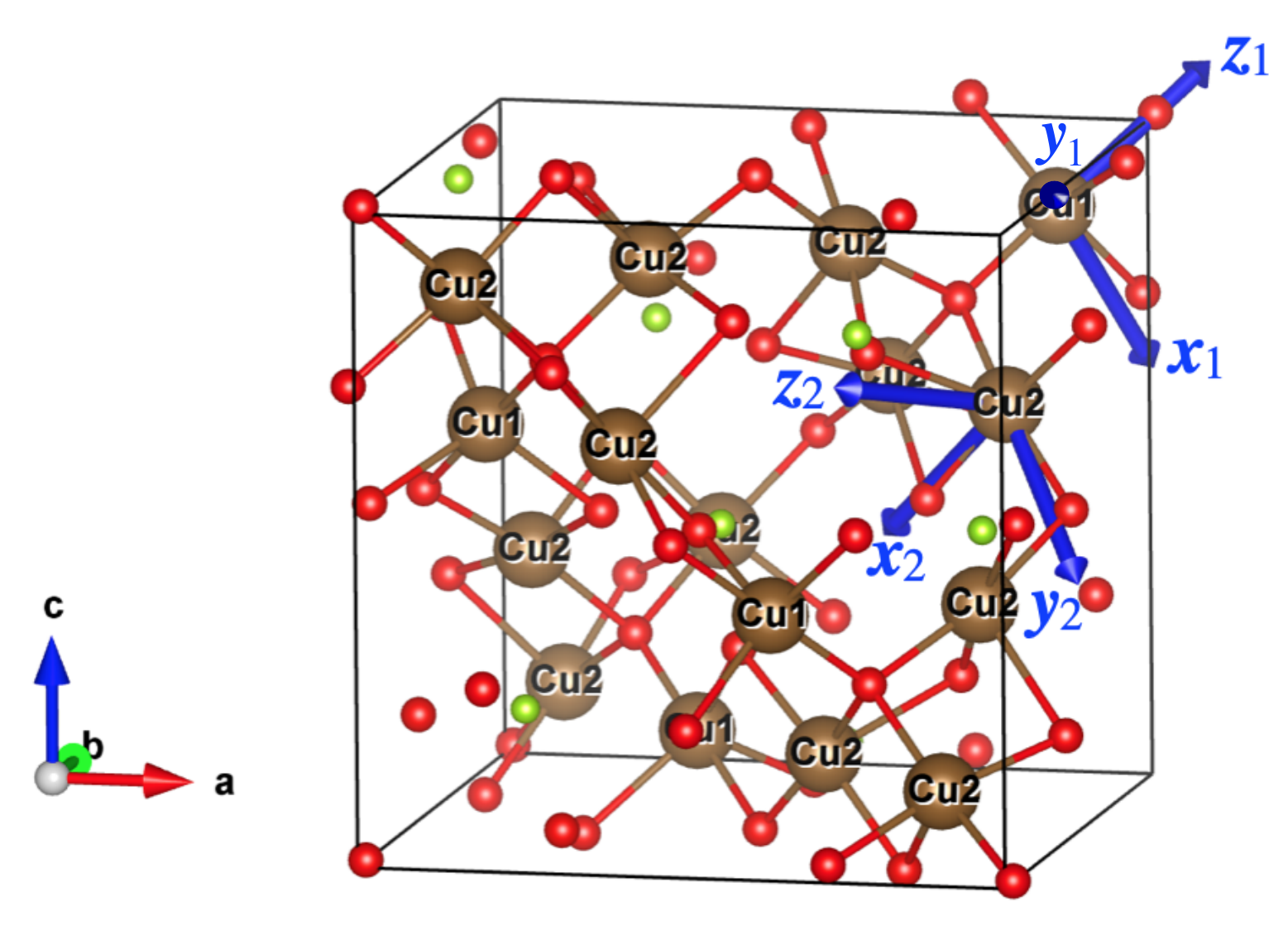}
\caption{The local coordindates (blue arrows) with respect to which the Wannier functions are defined. 
The local $z$-axis are chosen to be along the local 3-fold rotational axis for
the Cu-I sites ($z_1$) and normal to the CuO$_4$ plane for the Cu-II sites ($z_2$), respectively.}
\label{fig:localaxis}
\end{figure}

The $3d$ orbitals are defined with respect to local cartesian-coordinates shown in Fig.~\ref{fig:localaxis}, as well as in Fig.~2{\bf b} of the main text. 
The local $z$-axes are chosen to be along the local 3-fold rotational axis for the Cu-I sites and normal to the CuO$_4$ plane for the Cu-II sites, respectively. As a result of the crystallization environment, Cu-I with an approximately trigonal bipyramidal coordination, has a formal C$_3$ site symmetry. Cu-II instead has an approximately square pyramidal surrounding and a formal C$_1$ site symmetry, i.e. no symmetry. 

A tight binding (TB) Hamiltonian consisting of $3d$ orbitals from four Cu-I sites and twelve Cu-II sites is then obtained by the maximally localized Wannier functions method~\cite{marzari_maximally_2012, mostofi_wannier90_2008}. The local on-site crystal field Hamiltonian for the Cu-I and Cu-II sites are (energy in eV):



\begin{equation}
\hat{H}^{\text{CF}}_{\text{Cu-I}} =\bordermatrix{
                        &d_{z^2} &d_{xz} &d_{yz} & d_{x^2-y^2} & d_{xy}\cr
                d_{z^2} & -0.001  &  0         &  0         &  0         &  0         \cr       
                d_{xz}  &  0         &  -1.313  &  0         &  0.120  &  0         \cr
                d_{yz}  &  0         &  0         &  -1.313  &  0         &  -0.120 \cr
            d_{x^2-y^2} &  0         &  0.120  &  0         &  -1.178  &  0         \cr
                d_{xy}  &  0         &  0         &  -0.120 &  0         &  -1.178 
}, 
\label{eqn:cf_cu1}
\end{equation}

\begin{equation}
\hat{H}^{\text{CF}}_{\text{Cu-II}} =\bordermatrix{
                        &d_{z^2} &d_{xz} &d_{yz} & d_{x^2-y^2} & d_{xy}\cr
                d_{z^2} & -1.077  &   -0.009 &  -0.112 & 0.017 &  -0.020\cr  
                d_{xz}  & -0.009 &   -1.243 &  0.035 & -0.026 &  0.043\cr
                d_{yz}  & -0.112 &   0.035 &  -1.364 & -0.014 &  -0.011\cr
            d_{x^2-y^2} & 0.017 &   -0.026 &  -0.014 & 0.282  &  0.258\cr
                d_{xy}  & -0.020 &   0.043 &  -0.011 & 0.258 &  -1.077 
}. 
\hspace{1.5cm} \label{eqn:cf_cu2}
\end{equation}
\vspace{5mm}

\begin{figure}[h]
\includegraphics[width=0.85\textwidth]{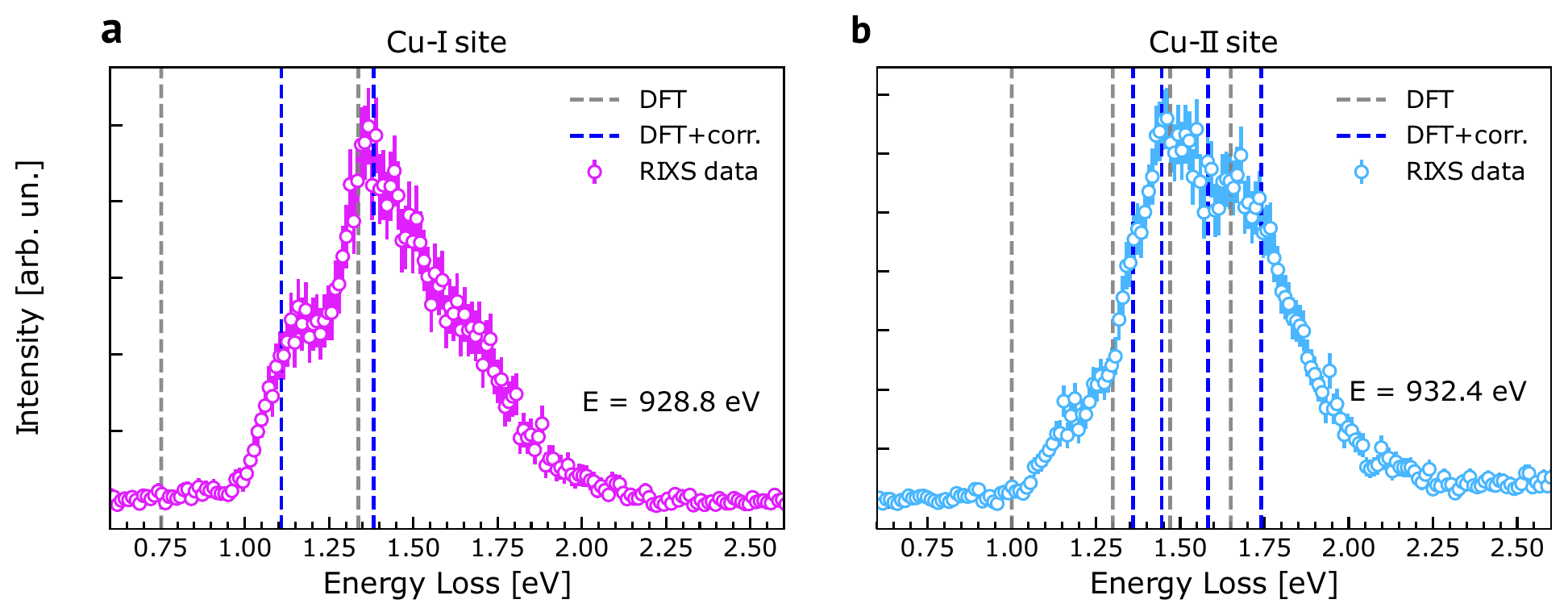}
\caption{\textcolor{black}{DFT levels for Cu-I site ({\bf a}) and Cu-II site ({\bf b}): raw DFT levels are displayed as grey dashed lines, while the corrected DFT levels resulting from Eqs. \ref{eqn:cf_cu1}-\ref{eqn:cf_cu2} are displayed as blue dashed lines. The calculated levels are overlaid to the RIXS spectra measured at E=928.8 eV (932.4 eV), mostly sensitive to Cu-I (Cu-II), respectively.}}
\label{fig:Fig_S_dd_DFT}
\end{figure}

Overall, the energy levels displayed in the above local crystal field Hamiltonians are slightly adjusted with respect to the original DFT energy levels, in order to obtain a \textcolor{black}{reasonable set of parameters for initializing the fitting of} the experimental RIXS results. This adjustment is necessary because the DFT calculations are not very accurate for the 3$d$ transition metal oxides. \textcolor{black}{This statement can be better understood referring to Fig. \ref{fig:Fig_S_dd_DFT}{\bf a} ({\bf b}), that  displays the RIXS spectrum measured before (after) the Cu L$_3$ edge as being mostly sensitive to Cu-I (Cu-II) site, respectively. The raw DFT levels, overlaid to the experimental data as grey dashed lines, clearly underestimate the crystal field energies for both Cu-I and Cu-II by as much as $\sim 0.45$~eV. Rather, the corrected DFT levels displayed as blue dashed lines in the same figure match well the experimental data. To be specific, the corrected DFT levels have been obtained by shifting for Cu-I the orbitals $d_{x^2-y^2}$ and $d_{xy}$ down by 0.4 eV; while for Cu-II the orbital $d_{z^2}$ down by 0.12 eV, the orbital $d_{x^2-y^2}$ up by 0.25 eV and the orbital $d_{xy}$ up by 0.45 eV.}

Since for C$_3$ site symmetry the five $3d$ orbitals split into one $A_{1g}$ representation ($d_{z^2}$) and two coinciding $E_g$ representations [($d_{xz}$, $d_{yz}$) and ($d_{x^2-y^2}$, $d_{xy}$)], the resulting $\hat{H}^{\text{CF}}_{\text{Cu-I}}$ in Eq. \ref{eqn:cf_cu1} is unavoidably a mix between the two $E_g$ representations, i.e. $d_{xz}$ mixes with $d_{x^2-y^2}$ and $d_{yz}$ mixes with $d_{xy}$. Due to the lack of symmetry for Cu-II site,  the terms in $\hat{H}^{\text{CF}}_{\text{Cu-II}}$ result completely mixed.

Diagonalizing $\hat{H}^{\text{CF}}_{\text{Cu-I}}$ and $\hat{H}^{\text{CF}}_{\text{Cu-II}}$ yields new  wavefunctions with the expected mixed orbital characters, as displayed in Tab. \ref{tab:tab4} for Cu-I and in Tab. \ref{tab:tab5} for Cu-II. For simplicity, we label the new wavefunctions as $d_{z^2}$, $d_{xz}$, $d_{yz}$, $d_{x^2-y^2}$, $d_{xy}$ by simply referring to their dominant orbital component (above 80\%), higlighted in blue in Tabs. \ref{tab:tab4}-\ref{tab:tab5}. For example, for Cu-I, the first wavefunction Wf1 = -0.863$d_{xz}$+0.505$d_{x^2-y^2}$ is referred to as $d_{xz}$.

The eigenvalues of the diagonalized  $\hat{H}^{\text{CF}}_{\text{Cu-I}}$ and $\hat {H}^{\text{CF}}_{\text{Cu-II}}$ provide the energy of the crystal field levels for Cu-I and Cu-II sites, respectively. These values are summarized in Tab. \ref{tab:tab2}, second and fifth columns, and are labelled by their dominant orbital character. The Cu-I topmost hole orbital energy (ground state, GS) is of reference energy and set to 0 eV. The GS level of the Cu-II sites is higher in energy than Cu-I by about 0.33 eV, so the Cu-II sites resonate at higher energy than the Cu-I sites.

Furthermore, for each Cu species, the energy difference between the ground state level with respect to each crystal field level is calculated in the third and sixth columns of Tab. \ref{tab:tab2}, respectively for Cu-I and Cu-II. Such values provide a good match with the $dd$-excitation energies extracted from the experiment, see Tab. \ref{tab:tab3}, and discussed in the main text. 

\begin{table}[h!]
    \centering
    \begin{tabular}{| c | c | c | c | c | c |}
    \hline
   {} &  {Wf1 [-1.383 eV]} &  {Wf2 [-1.383 eV]}  & {Wf3 [-1.108 eV]} &  {Wf4 [-1.108 eV]} &  {Wf5 [0 eV]} \\ [0.6ex]
    \hline
    {\textbf{$d_{z^2}$} } &  {0} & {0} & {0} & {0} & \textcolor{blue}{1}  \\[0.6ex]
    \hline
    {\textbf{$d_{xz}$}} & \textcolor{blue}{-0.863} & {0} & {0.505} &{0} &{0}  \\[0.6ex]
    \hline
    {\textbf{$d_{yz}$}}  & {0} & \textcolor{blue}{-0.863} & {0} &{-0.505}  &{0} \\[0.6ex]
    \hline
    {\textbf{$d_{x^2-y^2}$}} & {0.505} & {0} & \textcolor{blue}{0.863} & {0}  & {0} \\[0.6ex]
    \hline
    \textbf{$d_{xy}$}  & {0} & {-0.505} & {0} & \textcolor{blue}{0.863} & {0} \\  [0.6ex] 
    \hline
    \end{tabular}
    \caption{Orbital components of the Cu-I wavefunctions after diagonalization of $\hat{H}^{\text{CF}}_{\text{Cu-I}}$. The wavefunctions are progressively labelled as Wf1 to Wf5, in order of increasing eigenvalues.}
    \label{tab:tab4}
\end{table}

\begin{table}[h!]
    \centering
    \begin{tabular}{| c | c | c | c | c | c |}
    \hline
   {} &  {Wf1 [-1.410 eV]} &  {Wf2 [-1.253 eV]}  & {Wf3 [-1.115 eV]} &  {Wf4 [-1.030 eV]} &  {Wf5 [0.330 eV]} \\ [0.6ex]
    \hline
    {\textbf{$d_{z^2}$} } &  {0.311} & {-0.123} & {-0.269} & \textcolor{blue}{0.903} & {-0.009}  \\[0.6ex]
    \hline
    {\textbf{$d_{xz}$}} & {-0.196} & \textcolor{blue}{-0.927} & {-0.287} & {-0.144} & {0.012}  \\[0.6ex]
    \hline
    {\textbf{$d_{yz}$}}  & \textcolor{blue}{0.926} & {-0.182} & {0.113} &{-0.310}  &{0.010} \\[0.6ex]
    \hline
    {\textbf{$d_{x^2-y^2}$}} & {-0.012} & {-0.066} & {0.165} & {0.034}  & \textcolor{blue}{-0.984} \\[0.6ex]
    \hline
    \textbf{$d_{xy}$}  & {0.084} & {0.298} & \textcolor{blue}{-0.898} & {-0.257} & {-0.180} \\  [0.6ex] 
    \hline
    \end{tabular}
    \caption{Orbital components of the Cu-II wavefunctions after diagonalization of $\hat {H}^{\text{CF}}_{\text{Cu-II}}$. The wavefunctions are progressively labelled as Wf1 to Wf5, in order of increasing eigenvalues.}
    \label{tab:tab5}
\end{table}


\begin{table}[h!]
    \centering
    \begin{tabular}{| c | c | c | c | c | c |}
    \hline
    \multicolumn{3}{| c |}{\textbf{Cu-I}}  & \multicolumn{3}{|c|}{\textbf{Cu-II}}\\ [0.7ex]
    \hline
    {Orbital} &  {Energy [eV]} &  {$dd$-exc [eV]} & {Orbital} &  {Energy [eV]} & {$dd$-exc [eV]} \\ [0.6ex]
    \hline
    \textbf{$d_{z^2}$} (GS) &  0 &  0 & \textbf{$d_{x^2-y^2}$} (GS) &  0.330 &  0\\[0.6ex]
    \textbf{$d_{xy}$} &  -1.108 &  1.108 & \textbf{$d_{z^2}$} &  -1.030 &  1.360 \\[0.6ex]
    \textbf{$d_{x^2-y^2}$} &  -1.108 &  1.108 & \textbf{$d_{xy}$} &  -1.115 &  1.445\\[0.6ex]
    \textbf{$d_{xz}$} &  -1.383 &  1.383  & \textbf{$d_{xz}$} &  -1.253  &  1.583\\ [0.6ex]  
    \textbf{$d_{yz}$} &  -1.383 &  1.383 & \textbf{$d_{yz}$} &  -1.410  &  1.740 \\  [0.6ex] 
    \hline
    \end{tabular}
    \caption{Energy of the Cu-I and Cu-II calculated crystal-field levels and corresponding $dd$-excitation energies.}
    \label{tab:tab2}
\end{table}

\section{XAS and RIXS simulations at Cu-$L_3$ edge}

Using the crystal-field energies and the orbital characters identified within the above DFT section, we simulate the XAS and RIXS spectra based on single-atom model using the EDRIXS code~\cite{wang_edrixs_2019}, by accounting for all the differently oriented Cu ions within the unit cell. For the current case with only one hole per site, the spectra are not sensitive to the exact Slater parameters, so standard values and rescaling were used here \cite{cowan_theory_1981}. 

The selected geometrical configuration reproduces the experimental conditions. In particular, the scattering plane contains the [001] and [110] directions. X-rays with $\pi$-polarization are incident at 20$^{\circ}$ with respect to the [001] direction and the sum of the $\pi$ and $\sigma$ polarized x-rays scattered around 90$^{\circ}$ are analyzed. Moreover, the life times of intermediate and final states are 0.6 eV and 0.2 eV, respectively.

\begin{figure}[!ht]
\includegraphics[width=0.85\textwidth]{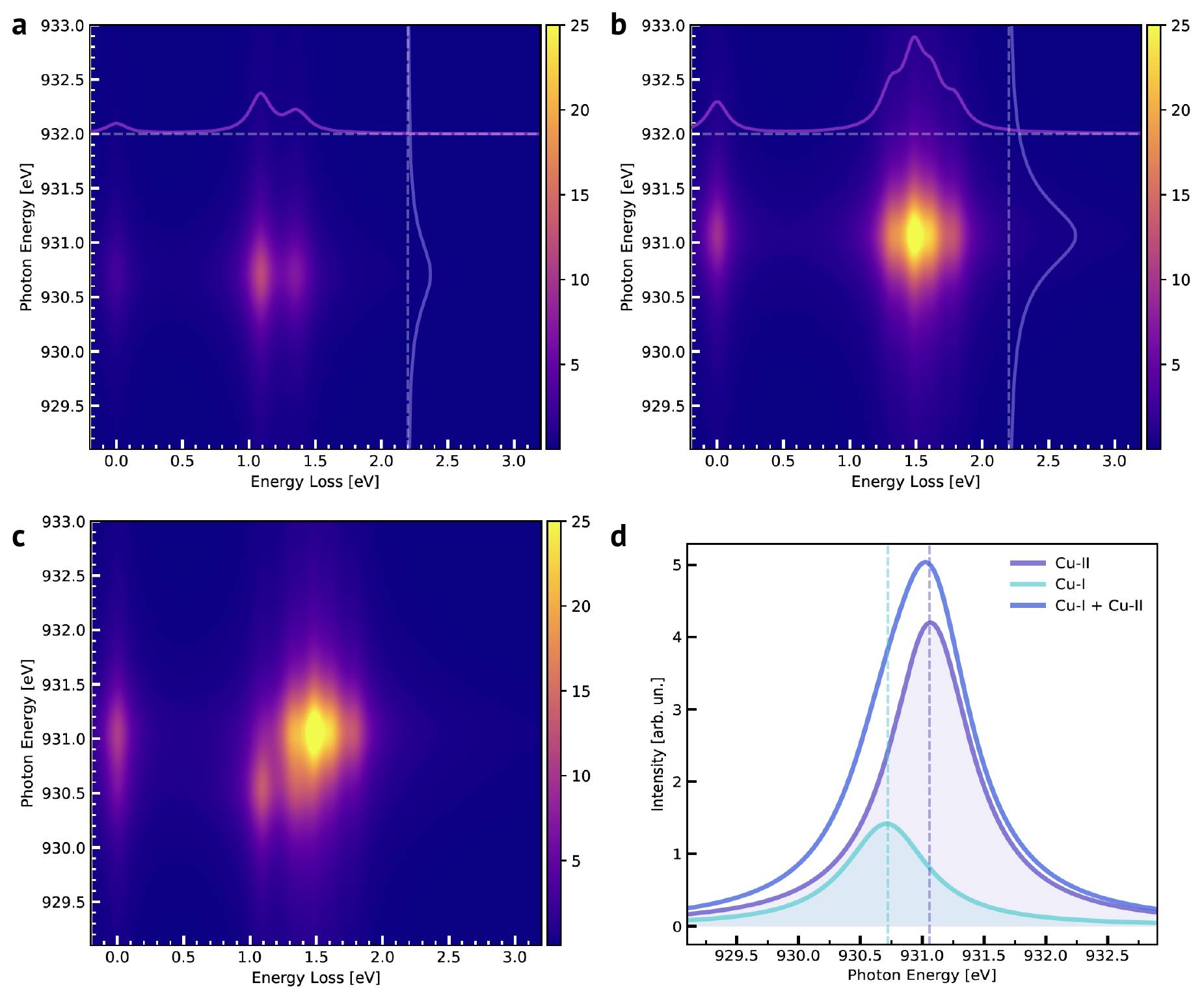}
\caption{Calculated RIXS map of {\bf a} the four Cu-I sites and {\bf b} the twelve Cu-II sites. {\bf c} The total RIXS map, contributed from the Cu-I and Cu-II sites. {\bf d} The integrated RIXS intensities 
as a function of the incident photon energy, as extracted from the calculated site responses and their sum.}
\label{fig:rixs}
\end{figure}

Fig.~\ref{fig:rixs}{\bf a} (~\ref{fig:rixs}{\bf b}) are the calculated RIXS map of the four Cu-I [twelve Cu-II] sites. There are two excitations, E1 and E3, for the Cu-I sites; and four
excitations, E2, E4, E5, E6, for the Cu-II sites. The resonant energies for Cu-I and Cu-II are both offset by the same value, $\sim931$ eV, to match the experimental finding. For each RIXS map, we perform the integral over the energy loss axis in order to obtain the corresponding absorption spectrum associated with each Cu site (see solid line in slate-blue color). The offset between the absorption maxima is determined by the energy difference between the Cu-I and Cu-II ground state levels. By performing instead the integral over the photon energy axis, we obtain the fundamental RIXS spectrum for each individual Cu site (see solid line in purple color). The fundamental RIXS spectrum provides  information on the 
$dd$-excitations cross-section for each site. These ratios are then used for the analysis of the experimental data. The sum of the Cu-I and Cu-II RIXS map represents the total RIXS map [see Fig.~\ref{fig:rixs}{\bf c}], which can be compared with the experimental data shown in Fig. 1{\bf c} of the main text.
Fig.~\ref{fig:rixs}{\bf d} instead reproduces the individual absorption spectra obtained for each Cu sites and their sum, which provides the total absorption spectrum in the approximation of no interference.

\section{Fitting of the RIXS data to disentangle the site-dependent $dd$-excitations}
The Cu $L_3$ RIXS spectrum of Cu$_2$OSeO$_3$ is dominated by crystal-field $dd$-excitations in the 1-2 eV energy range (see Fig. 2{\bf a} of the main text). The energy-proximity of the two Cu species present in this material  means that 
a theory-constrained fit analysis is required. 

By using the DFT energies for the $dd$-excitations in Tab. \ref{tab:tab2}, and the RIXS intensity obtained from the single-atom model calculations (see Sec. III), we can simulate the fundamental RIXS spectra associated with each Cu species, as displayed in Fig.2{\bf c} for Cu-I and in Fig.2{\bf d} for Cu-II by using equal-width Gaussian profiles for each $dd$ excitation peak. 
These fundamental spectra allow to extract the following intensity ratios: $I_{E1}:I_{E3} = 1: 1.23$ for Cu-I and $I_{E2}:I_{E4}: I_{E5}: I_{E6}= 1: 1.62:1.16:1.05$ for Cu-II, where the lowest-energy $dd$-excitation intensity for each Cu species is taken as a reference. These ratios are useful for fitting the data, when different widths for the $dd$-excitations could be used.

Next, we fit the RIXS spectra displayed in Fig.~2{\bf a}. By using a normalized Gaussian profile for each $dd$-excitation, we build the following fitting model:
\begin{equation}
    I_\text{RIXS}^\text{total}=I_\text{RIXS}^{\text {CuI}}+I_\text{RIXS}^{\text {CuII}}
    \label{eqn:fit1}
\end{equation}
where,

\begin{equation}
    I_\text{RIXS}^{\text {CuI}} = A^{\text{CuI}} \cdot (G_1+1.23 \cdot G_3)
    \label{eqn:fit2}
\end{equation}

\begin{equation}
    I_\text{RIXS}^{\text {CuII}} = A^{\text{CuII}} \cdot (G_2+1.62 \cdot G_4+1.16 \cdot G_5+1.05 \cdot G_6)
    \label{eqn:fit3}
\end{equation}

with $G_1$-$G_6$ being area-normalized Gaussian functions, and $A^{\text{CuI}}$/$A^{\text{CuII}}$ being amplitude factors for Cu-I/Cu-II site, respectively.
In reference to the area-normalized RIXS spectra mentioned in the main text, $(G_1+1.23 \cdot G_3)$ = $\tilde{I}_{RIXS}^{\text {CuI}}$, with a total area of 2.23, while $(G_2+1.62 \cdot G_4+1.16 \cdot G_5+1.05 \cdot G_6)$ = $\tilde{I}_{RIXS}^{\text {CuII}}$, with a total area of 4.83.

To perform the fitting of the RIXS spectra versus incident energy, we leave the amplitude parameters $A^{\text{CuI}}$/$A^{\text{CuII}}$ completely free, as well as the width of the Gaussian profiles. Instead, we constrain: {\textit i)} the Gaussian centers within $\pm$ 175 meV from the DFT $dd$-excitation energies reported in Tab. \ref{tab:tab2}; {\textit ii)} the $dd$-excitations intensity ratio to the single-ion calculated values as set in the fitting model.

The fitted spectra are presented in Fig.~3{\bf a-r}, where each panel refers to a specific incident energy and displays the raw data (circular markers), the fitted Gaussian components $G_1$ to $G_6$ (solid filling) and their sum (solid line).

The first main result of the fitting is the center position of the $G_1$ - $G_6$ components, reported in Fig. ~3{\bf s-t}. The values display very small variation as a function of incident energy, indicating that the fitting model is pretty adequate. Their average and error (as standard deviation versus the values obtained for each incident energy) are summarized in Tab. \ref{tab:tab3} (and in the fourth column of Tab. 2). Furthermore, the averaged energies are reasonably close to the calculated values (see Tab. \ref{tab:tab2}). The maximum deviation between the experimental and calculated values is $\sim$ 120 meV, well below the $\pm$ 175 meV value set in the fitting constraint, confirming that this choice is appropriate.

The averaged full width half maximum (FWHM) of the fitted Gaussians as a function of incident energy is presented in the fourth column of Tab. \ref{tab:tab3}, together with the respective error (again, considered as standard deviation). 

\begin{table}[h]
    \centering
    \begin{tabular}{|c|c|c|c|}
        \hline
        {\bf{Site}} & {\bf{Excitation}} & {\bf{Exp. Energy [eV]}}  & {\bf{Exp. FWHM [eV]}} \\[.5ex]
        \hline
        \multirow{2}{*}{Cu-I}  & \multicolumn{1}{c|}{E1} &   \multicolumn{1}{c|}{1.18 $\pm$ 0.015} &   \multicolumn{1}{c|}{0.19 $\pm$ 0.01} \\[.6ex]
        & \multicolumn{1}{c|}{E3} &  \multicolumn{1}{c|}{1.40 $\pm$ 0.010} &  \multicolumn{1}{c|}{0.21 $\pm$ 0.012} \\[.6ex]
        \hline
        \multirow{4}{*}{Cu-II}  & \multicolumn{1}{c|}{E2} &  \multicolumn{1}{c|}{1.40 $\pm$ 0.010} &  \multicolumn{1}{c|}{0.21 $\pm$ 0.012}\\[.6ex]
        & \multicolumn{1}{c|}{E4} &  \multicolumn{1}{c|}{1.57 $\pm$ 0.010} &  \multicolumn{1}{c|}{0.26 $\pm$ 0.017}\\[.6ex]
        & \multicolumn{1}{c|}{E5}  &  \multicolumn{1}{c|}{1.67 $\pm$ 0.012} &  \multicolumn{1}{c|}{0.27 $\pm$ 0.018}\\[.6ex]
        & \multicolumn{1}{c|}{E6} &  \multicolumn{1}{c|}{1.86 $\pm$ 0.014} &  \multicolumn{1}{c|}{0.31 $\pm$ 0.020}\\[.6ex]
        \hline
    
    \end{tabular}
    \caption{Fit results averaged over for all the RIXS spectra measured across the Cu L$_3$ edge: the first column reports the label assigned to the $dd$-excitations according to Fig. 2{\bf b}; the second columns displays the center values for the $dd$-excitations; and the third column displays the FWHM values.}
    \label{tab:tab3}
\end{table}

Finally, we can note that the FWHM of the $dd$-excitations systematically increases with the increasing of the $dd$-excitation energy itself: this can be explained with a shortening of the lifetime associated with the higher-energy of the excitations.

\makeatletter
\renewcommand\@biblabel[1]{#1.}
\makeatother

\bibliographystyle{naturemag}
\bibliography{aps}